\def\sqr#1#2{{\vcenter{\vbox{\hrule height.#2pt\hbox{\vrule width.#2pt
height#1pt \kern#1pt \vrule width.#2pt}\hrule height.#2pt}}}}

\def\fii{\varphi}

\def\=d{\,{\buildrel\rm def\over =}\,}

\def\i3p{\p32\int d^3p}

\def\As{A\hbox to 1pt{\hss /}}
\def\np4{\int d^4p_1\cdots d^4p_{n-1}\, }

\def\supp{{\rm supp}\, }

\def\sgn{{\rm sgn}\, }

\def\iinf{\int\limits_{-\infty}^{+\infty}}

\def\nx4{\int d^4x_1\ldots d^4x_n\, }

\def\kon#1#2{\vbox{\halign{##&&##\cr
\lower4pt\hbox{$\scriptscriptstyle\vert$}\hrulefill &
\hrulefill\lower4pt\hbox{$\scriptscriptstyle\vert$}\cr $#1$&
$#2$\cr}}}
\def\lra{\longleftrightarrow}
\def\konv#1#2#3{\hbox{\vrule height12pt depth-1pt}
\vbox{\hrule height12pt width#1cm depth-11.6pt}
\hbox{\vrule height6.5pt depth-0.5pt}
\vbox{\hrule height11pt width#2cm depth-10.6pt\kern5pt
      \hrule height6.5pt width#2cm depth-6.1pt}
\hbox{\vrule height12pt depth-1pt}
\vbox{\hrule height6.5pt width#3cm depth-6.1pt}
\hbox{\vrule height6.5pt depth-0.5pt}}
\def\konu#1#2#3{\hbox{\vrule height12pt depth-1pt}
\vbox{\hrule height1pt width#1cm depth-0.6pt}
\hbox{\vrule height12pt depth-6.5pt}
\vbox{\hrule height6pt width#2cm depth-5.6pt\kern5pt
      \hrule height1pt width#2cm depth-0.6pt}
\hbox{\vrule height12pt depth-6.5pt}
\vbox{\hrule height1pt width#3cm depth-0.6pt}
\hbox{\vrule height12pt depth-1pt}}

\def\konw#1#2#3{\hbox{\vrule height12pt depth-1pt}
\vbox{\hrule height12pt width#1cm depth-11.6pt}
\hbox{\vrule height6.5pt depth-0.5pt}
\vbox{\hrule height12pt width#2cm depth-11.6pt \kern5pt
      \hrule height6.5pt width#2cm depth-6.1pt}
\hbox{\vrule height6.5pt depth-0.5pt}
\vbox{\hrule height12pt width#3cm depth-11.6pt}
\hbox{\vrule height12pt depth-1pt}}

\def\i{{\rm int}}

\def\m3{{\mu_1\mu_2\mu_3}}

\def\p{{(+)}}

\documentstyle[11pt]{article}

\columnwidth\textwidth
\title{The Two-Loop Master Diagram in the Causal Approach \thanks{Work
supported by Swiss National Science Foundation}}
\author{A.Aste and G.Scharf\\
Institut f\"ur Theoretische Physik der Universit\"at Z\"urich\\
Winterthurerstrasse 190, CH-8057 Z\"urich, Switzerland}
\date{\today}
\begin{document}
\maketitle

\begin{abstract}
The scalar two-loop master diagram is revisited in the massive cases needed
for the computation of boson and fermion propagators in QED and QCD. By means
of the causal method it is possible in a straightforward manner to express
the propagators as double integrals. In the case of vacuum polarization both
integrations can be carried out in terms of polylogarithms, whereas the last
integral in the fermion propagator cannot be expressed by known special
functions. The advantage of the method in comparison with Feynman integral
calculations is indicated.
\\
\\

interactions.
\end{abstract}
\vspace{-12 cm}
\centerline{Preprint:ZU-TH-22/96, hep-th/9608193}
\vspace{12 cm}
\newpage
\section{Introduction}
It is the big advantage of Epstein and Glaser's
causal approach to quantum field theory [1]
that it explicitly uses the causal structure of the theory, which is
invisible in the traditional Lagrangian approach using Feynman rules or
in path-integral methods. As a consequence, it is possible to prove the
general properties of the S-matrix in very direct and transparent manner [2].
However, the full strength of the causal method comes out in higher order
calculations and can be summarized in the following points:

1.) There exists no ultraviolet problem in the causal theory, so that no
regularization is necessary.

2.) All integrals are four-dimensional, so that no $\gamma^5$-problem
exists. Furthermore, the number of non-trivial integrations is reduced
to the minimum, essentially one one-dimensional integral for every additional
loop.

3.) The method is inductive in the order $n$ of perturbation theory, so
that the work done in lower orders is completely and automatically
utilized in higher orders.

The main reason why the Feynman rules are not optimal for loop diagrams is
the following. Suppose that time ordering can simply be done by multiplication
with step functions $\Theta (x_k^0 - x_j^0)$, as in the scalar theory we
are going to consider. Then, in calculating a time-ordered product according
to the rules, {\bf every} pairing of field operators gets such a
$\Theta$-function  which leads to the Feynman propagators $D_F (x_k - x_j)$.
However, the time ordering of the vertices is already specified by less
$\Theta$-functions. Without the temporal $\Theta$-function the pairing function
is $D^+$ instead of $D_F$. Consequently, some propagators $D_F$ can actually
be simplified into $D^{\pm}$ which is simpler to integrate because it contains
a $\delta$-distribution in p-space. For example, in the causal theory the
third order scalar vertex function is given by
$$
\Lambda(x_1,x_2,x_3) = D_F (x_1-x_3) D^+ (x_3-x_2) D_F (x_1-x_2)
-D_F D^{av} D^- + D^+ D^{av} D^{ret} . \eqno{(1.1)}
$$
If one substitutes $D^+=D_F - D^{av}$ in the first term, one arrives
at the usual Feynman form
$$
\Lambda = D_F D_F D_F - D_F D^{av} (D_F+D^-) + D^+ D^{av} D^{ret}, \eqno{(1.2)}
$$
because the last two terms are equal to $D^{av} D^{av} D^{ret}$ which
vanishes by the support properties of the advanced and retarded distributions.
But (1.1) is simpler to calculate
than (1.2) because every member contains one $D^\pm$.
This advantage is strongly increasing in higher loop diagrams.

It is the purpose of this paper to illustrate these features for the
so-called two-loop master diagram of Fig.1. Interestingly enough, this
diagram was already computed with the causal method by various authors
without knowing it. The first were K\"all\'en and Sabry [3], their
work was extended by Broadhurst [4] to other mass cases. The most
extensive higher order calculations using the "dispersive method" were
carried out over many years by the italian group Mignaco, Remiddi [5],
Barbieri [6] and others (see [7] and references given therein).
All these authors base
their calculations on analytic properties of Feynman integrals which are
often referred to as Cutcosky rules. The lack of understanding of these
analytic
properties has created the problem of anomalous thresholds. Fortunately,
the latter do not appear in diagrams with two and three external legs,
but they do appear in four- and more legs-diagrams [8].

Such basic subjects are discussed in the following section. As a first
application, we then briefly describe the calculation of the scalar vertex
function with arbitrary masses which is needed at various places in the
later two-loop calculations. In Sect.3 we turn to the master diagram for
vacuum polarization. We show the calculation of the causal distribution in
some detail to localize the infrared divergences which appear in the case of
vanishing photon mass. The most difficult integration is the dispersion
integral for the splitting. The necessary techniques to handle polylogarithms
are shown in Sect.4. Finally we consider the more complicated case of the
fermion propagator. By ingenious tricks, Broadhurst [4] succeeded in expressing
this propagator by a one-dimensional integral over the complete elliptic
integral of the first kind $\mbox{K} (k)$ together with elementary functions
which can easily be computed numerically. In the causal theory one needs not be
a genius. We get the result by a slight change of the standard procedure.
As far as the spin structure is concerned,
it is known that the general two-legs diagram can be expressed in terms
of the scalar two-point functions [9]. Therefore, we restrict ourselves
to the scalar case here.

\vskip 1cm
\section{The Causal Method and the Scalar Vertex}
\vskip 1cm
We first summarize the main ingredients of the causal method, for
details we refer to [1,2]. In the causal theory the S-matrix is viewed as an
operator-valued distribution of the following form
$$S(g)=1+\sum_{n=1}^\infty{1\over n!}\int dx_1\ldots dx_n\,T_n(x_1,
\ldots x_n)g(x_1)\cdot\ldots g(x_n),\eqno(2.1)$$
where $g\in S({\bf R^4})$, the Schwartz space of functions of rapid
decrease. The test function
$g$ plays the role of "adiabatic switching" and provides a cutoff in the
long-range part of the interaction, without destroying any symmetry. It
can be considered as a natural infrared regulator. The adiabatic limit
$g\to 1$ must be performed at the end of the calculation in the right
quantities where this limit exists. The existence of the adiabatic limit
becomes a problem if the theory contains massless fields.

The $n$-point operator-valued distributions $T_n$ are the basic objects
of the theory. They can be constructed inductively from $T_1$ through a
number of physical requirements, the most essential one being causality.
Unitarity plays no essential role.
Let the operator-valued distributions $\tilde T_n$ be defined by
$$S(g)^{-1}=1+\sum_{n=1}^\infty{1\over n!}\int d^3x_1\ldots d^3x_n\,
\tilde T_n(x_1,\ldots x_n)g(x_1)\ldots g(x_n).\eqno(2.2)$$
Then one defines, for arbitrary sets of points $X$, $Y$ in
Minkowski space, the following distributions
$$A'_n(x_1,\ldots x_n)=\sum_{P_2}\tilde T_{n_1}(X)T_{n-n_1}(Y,x_n)\eqno
(2.3)$$
$$R'_n(x_1,\ldots x_n)=\sum_{P_2}T_{n-n_1}(Y,x_n)\tilde T_{n_1}(X),
\eqno(2.4)$$
where the sums run over all partitions
$$P_2:\quad\{x_1,\ldots x_{n-1}\}=X\cup Y,\quad X\ne\emptyset$$
into disjoint subsets with $|X|=n_1$, $|Y|\le n-2$, and $|X|$ means the
number of points in the set $X$. We also introduce
$$D_n(x_1,\ldots x_n)=R'_n-A'_n.\eqno(2.5)$$
If the sums are extended over all partitions $P_2^0$, including the
empty set $X=\emptyset$, we obtain the distributions
$$A_n(x_1,\ldots x_n)=\sum_{P_2^0}\tilde T_{n_1}(X)T_{n-n_1}(Y,x_n)=$$
$$=A'_n+T_n(x_1,\ldots x_n),\eqno(2.6)$$
$$R_n(x_1,\ldots x_n)=\sum_{P_2^0}T_{n-n_1}(Y,x_n)\tilde T_{n_1}(X)=$$
$$=R'_n+T_n(x_1,\ldots x_n).\eqno(2.7)$$
These two distributions are not known by the induction assumption
because they contain the unknown $T_n$. Only the difference
$$D_n=R'_n-A'_n=R_n-A_n\eqno(2.8)$$
is known.

One can determine $R_n$ or $A_n$
separately by investigating the support properties of the various
distributions, this is the point where the causal structure becomes
important. It turns out that $R_n$ is a retarded and $A_n$ an
advanced distribution
$$\supp R_n\subseteq \bar \Gamma^+_{n-1}(x_n),\quad\supp A_n\subseteq
\bar \Gamma^-_{n-1}(x_n)\eqno(2.10)$$
with
$$ \bar \Gamma^\pm_{n-1}(x)\equiv\{(x_1,\ldots x_{n-1})\>|\>x_j\in\bar V^\pm
(x),\forall j=1,\ldots n-1\}\eqno(2.11)$$
$$\bar V^\pm(x)=\{y\>|\>(y-x)^2\ge 0,\>\pm (y^0-x^0)\ge 0\}.\eqno(2.12)$$
Hence, by splitting of the causal distribution (2.8) one gets $R_n$
(and $A_n$), and $T_n$ then follows from (2.7) (or (2.6)). The $T_n$'s
are well-defined time-ordered products.
To carry out the splitting process, we write (2.8) in normally ordered
form and then split the numerical distributions $d_n^k(x)$, where
$x=(x_1-x_n,...,x_{n-1}-x_n)$.
The causal splitting of $d(x)$ can directly be done in momentum space by
means of the following dispersion formula
$$\hat r(p)=\pm{i\over 2\pi}\iinf dt\,{\hat d(tp)\over (t\mp
i0)^{\omega+1}(1-t\pm i0)},\eqno(2.13)$$
which holds true for $p\in\Gamma^+$ (upper signs) or $p\in\Gamma^-$
(lower signs). The result for arbitrary $p$ is
obtained by analytic continuation, based on the fact that the retarded
distribution $\hat r(p)$ is the boundary value of an analytic function,
regular in ${\bf R^{4n-4}}+i\Gamma^+$.
This so-called central splitting solution is very convenient because
it is obviously Lorentz covariant and does not destroy any symmetry of
the theory. $\omega$ is the singular order of $\hat d$ [2], in case of
trivial splitting one has $\omega = -1$.

The formulae (2.3-4) contain what is usually called Cutkosky rules. But
we would like to emphasize that the sums run over {\it all} non-trivial
partitions $X\cup Y$. If $X$ and $Y$ are connected the term
corresponds to an ordinary cut through the diagram. Otherwise the
diagram is decomposed into more than two pieces. In this situation
anomalous thresholds appear. But this only occurs in
diagrams with more than three external legs.

To calculate scalar diagrams with arbitrary masses, we start the inductive
process from the following first order
$$T_1(x)=i:\fii_1^+(x)\fii_2(x):A(x)-{\rm h.c.}=-\tilde T_1(x).
\eqno(2.14)$$
Here $\fii_1$, $\fii_2$ are charged scalar fields with masses $m_1$ and
$m_2$, respectively, and $A(x)$ is a neutral (self-adjoint) scalar field
of mass $m_3$.
All fields are free fields satisfying the following commutation relations
$$[\fii_j(x),\fii_j^+(y)]=-iD_{m_j}(x-y),\quad j=1,2\eqno(2.15)$$
$$[\fii_j^{(-)}(x),\fii_j^{+(+)}(y)]=-iD_{m_j}^{+}(x-y),\eqno(2.16)$$
where $D_m$ is the Jordan-Pauli distribution of mass $m$ and $(\pm)$
refers to the positive and negative frequency parts of the various quantities.
$A(x)$ fulfills the same commutation relations without the hermitian
adjoint $^+$. The commutator (2.16) gives the contraction in Wick's
theorem, its Fourier transform is equal to
$$\hat D^{+}_m(p)={i\over 2\pi}\Theta(p^0)\delta(p^2-m^2).\eqno(2.17)$$
The last vertex $x_n$ plays a special role in the above equations
(2.3-10). It is the splitting vertex, defining the edge of the causal
cone. By translation invariance the numerical distributions only depend
on the relative coordinates
$$y_j=x_j-x_n,\quad j=1,\ldots n-1.\eqno(2.18)$$
The Fourier transform is always understood with respect to these
relative coordinates
$$\hat d(p)=(2\pi)^{-2n+2}\int d(y)e^{ipy}\,d^4y_1\ldots d^4y_{n-1}.
\eqno(2.19)$$

Until now all $n$-th order distributions depend on $n$ or $n-1$
variables, the inner vertices are not integrated out. If the adiabatic
limit exists, we can integrate the inner coordinates with $g(x)=1$. In
$p$-space this means that the inner momenta are put equal to 0. Then
many terms in (2.3-4) vanish:

{\bf Lemma 1.} In the adiabatic limit only those partitions $X\cup Y$
contribute to $A'_n$ (2.3) where $X$ and $\{Y,x_n\}$ contain external
vertices, and similarly for $R'_n$ (2.4).

{\bf Proof.} Consider a partition where $X$ contains no external vertex.
Performing the contractions between $X$ and $\{Y,x_n\}$ with
$D^+$-distributions and transforming into $p$-space (2.17), we get a
product of $\Theta$-functions
$$\Theta(p_1^{\prime 0})\Theta(p_2^{\prime 0})\ldots\Theta(p_j^{\prime 0}),$$
where all momenta add up to 0:
$$p_1^{\prime 0}+p_2^{\prime 0}+\ldots +p_j^{\prime 0}=0.$$
Such a product is zero.

The exists one serious problem, however. The central splitting solution
(2.13) is only true if all momenta $p_j$ are {\bf inside} the light cone.
Therefore, strictly speaking, we cannot put the inner momenta equal to 0.
But if the $D_m^+$-distribution is massive $m>0$, the vanishing of the
contribution of a wrong partition takes place for small enough inner momenta
$\tilde p_j$ in $V^+$,already.
Then we can use (2.13) and take the limit $\tilde p_j
\rightarrow 0$. For this reason we always calculate with massive fields first.
If the limit $m \rightarrow 0$ is required, it must be carefully performed
by taking cancellations of infrared divergences between different terms
into account.

From lemma 1 it is clear that in diagrams with two and three external
legs, the non-vanishing terms (in the adiabatic limit) correspond to
ordinary cuts through the diagram. But in a four-legs diagram a pair of
opposite legs can be in $X$ and the other pair of opposite legs in
$\{Y,x_n\}$. Then this decomposition is no longer a simple cut and an
"anomalous threshold" appears. To our knowledge such a diagram was never
computed by the naive "dispersive method", but in the causal theory this
is no problem. The following second lemma further simplifies the later
calculations. It is a consequence of parity- and time-reversal invariance.

{\bf Lemma 2.} In a PT-invariant theory the numerical distributions
$d_n^k(x)$ in $D_n$ (2.5) are essentially real (i.e. up to an overall
factor $i$) in momentum space: $\hat d_n^k(p)^*=\hat d_n^k(p)$.

{\bf Proof.} We know that the causal $D$-distributions are PT-invariant
([2], p.281). This implies for the numerical distributions:
$d_n^k(-x)=d_n^k(x)^*$. The complex conjugate comes from the
antiunitarity of time-reversal. After Fourier transformation this gives
the desired result in momentum space. The overall factor $i$ depends on
whether the number of internal lines in the diagram is even or odd.

We now come to the calculation of the third order vertex diagram (Fig.2).
To simplify the notation, we write the arguments in $x$-space without the
dummy variable $x$. From (2.4) we have
$$R'_3=T_2(1,3)\tilde T_1(2)+T_2(2,3)\tilde T_1(1)+T_1(3)\tilde T_2(1,2).
\eqno(2.20)$$
The first term herein contains a Compton subgraph
$$R'_{31}=i:\fii_2^+(1)D_{m_1}^F(1-3)\fii_2(3)::A(1)A(3):(-i)A(2)
:\fii_2^+\fii_1(2):,\eqno(2.21)$$
where
$$\hat D_m^F(p)={-(2\pi)^{-2}\over p^2-m^2+i0}\eqno(2.22)$$
is the Feynman propagator. The product (2.21) is computed by Wicks
theorem, restricting ourselves to those contractions which generate the
vertex diagram:
$$R'_{31}=-:\fii_2^+(1)D_{m_1}^F(1-3)D_{m_2}^+(3-2)D_{m_3}^+(1-2)
\fii_1(2):A(3).\eqno(2.23)$$
Similarly we  get for the other two terms in (2.20)
$$R'_{32}=-:\fii_2^+(1)D_{m_1}^+(3-1)D_{m_2}^F(3-2)D_{m_3}^+(2-1)
\fii_1(2):A(3),\eqno(2.24)$$
$$R'_{33}=-:\fii_2^+(1)D_{m_1}^+(3-1)D_{m_2}^+(3-2)D_{m_3}^{AF}(1-2)
\fii_1(2):A(3).\eqno(2.25)$$
The anti-Feynman propagator $D^{AF}$ is the complex conjugate of $D^F$.
It appears because we have used unitarity $\tilde T_2(1,2)=T_2(1,2)^+$.
This is the only minor role which unitarity plays here.

The result for $A'_3$ (2.3) is obtained in the same way. Collecting the
terms with field operators $:\fii_2^+(1)\fii_1(2):A(3)$ in $D_3$ (2.5),
the corresponding numerical distribution is given by
$$d_3(1,2,3)=-D_{m_1}^F(1-3)D_{m_2}^+(3-2)D_{m_3}^+(1-2)-D^-_{m_1}
D^F_{m_2}D^-_{m_3}$$
$$+D^-_{m_1}D^+_{m_2}D^{AF}_{m_3}+D^F_{m_1}D^-_{m_2}D^-_{m_3}
+D^+_{m_1}D^F_{m_2}D^+_{m_3}-D^+_{m_1}D^-_{m_2}D^{AF}_{m_3},\eqno(2.26)$$
where we have used $D^-(x)=-D^+(-x)$ and the arguments in the 6 terms
agree with the first term. The 6 terms come from three cuts through the
vertex diagram, not only one. To get contact with the convention in [2],
we shall use the relative coordinates
$$y_1=x_1-x_3,\quad y_2=x_3-x_2,\eqno(2.27)$$
and calculate the Fourier transform
$$\hat d_3(p,q)=(2\pi)^{-4}\int d_3(y_1,y_2)e^{ipy_1+iqy_2}d^4y_1
d^4y_2.\eqno(2.28)$$
Then we arrive at
$$\hat d_3(p,q)=(2\pi)^{-2}\int dk\,\Bigl[D^-_{m_1}(p-k)D^+_{m_2}
(q-k)D^{AF}_{m_3}(k)$$
$$-D^F_{m_1}(p-k)D^+_{m_2}(q-k)D^+_{m_3}(k)-D^-_{m_1}(p-k)D^F_{m_2}
(q-k)D^-_{m_3}(k)\Bigl]-[p\lra q,\>m_1\lra m_2].\eqno(2.29)$$
Up to the arbitrary masses and the imaginary parts,
this agrees precisely with the result in
QED ([2], eq.(3.8.28)). By the same techniques as in the QED case,
we then find
$$\hat d_3(p,q)={\pi\over 4(2\pi)^6}\biggl\{{\sgn P_0\over\sqrt{N}}
\Theta(P^2-(m_1+m_2)^2)\log_1$$
$$-{\sgn q_0\over\sqrt{N}}\Theta(q^2-(m_2+m_3)^2)\log_2+{\sgn p_0
\over\sqrt{N}}\Theta(p^2-(m_1+m_3)^2)\log_3 \biggr\}, \eqno(2.30)$$
where $P=p-q, N=(pq)^2-p^2q^2$ and
$$\log_1=\log\Biggl|{p^2+m_1^2-m_3^2-pP\Bigl(1+{m_1^2-m_2^2\over P^2}\Bigl)
+\sqrt{N}\sqrt{1-2{m_1^2+m_2^2\over P^2}+({m_1^2-m_2^2\over P^2})^2}
\over p^2+m_1^2-m_3^2-pP\Bigl(1+{m_1^2-m_2^2\over P^2}\Bigl)
-\sqrt{N}\sqrt{1-2{m_1^2+m_2^2\over P^2}+({m_1^2-m_2^2\over P^2})^2}}\Biggr|,
\eqno(2.31)$$
$$\log_2=\log\Biggl|{p^2-m_1^2+m_3^2-pq\Bigl(1-{m_2^2-m_3^2\over q^2}\Bigl)
+\sqrt{N}\sqrt{(1-{m_2^2-m_3^2\over q^2})^2-{4m_3^2\over q^2}}\over
p^2-m_1^2+m_3^2-pq\Bigl(1-{m_2^2-m_3^2\over q^2}\Bigl)
-\sqrt{N}\sqrt{(1-{m_2^2-m_3^2\over q^2})^2-{4m_3^2\over q^2}}}\Biggr|,
\eqno(2.32)$$
$$\log_3=\log\Biggl|{q^2-m_2^2+m_3^2-pq\Bigl(1-{m_1^2-m_3^2\over p^2}\Bigl)
+\sqrt{N}\sqrt{(1-{m_1^2-m_3^2\over p^2})^2-{4m_3^2\over p^2}}\over
q^2-m_2^2+m_3^2-pq\Bigl(1-{m_1^2-m_3^2\over p^2}\Bigl)
-\sqrt{N}\sqrt{(1-{m_1^2-m_3^2\over p^2})^2-{4m_3^2\over p^2}}}\Biggr|.
\eqno(2.33)$$
Because of lemma 2 we only need the real parts of the logarithms.
The splitting of (2.30) by means of the central solution (2.13) is done
later in (3.10).

\section{Vacuum Polarization in Fourth Order}
Now we envisage the calculation of the diagram shown in Fig.1a,
which contributes to vacuum polarization in fourth order.
We restrict ourselves to the case with vanishing 'photon' mass $m_3$
and mass $m$ of the 'electron'.
From (2.4) we have
$$R'_4 = T_3 (1,2,4) \tilde T_1 (3) + T_3 (1,3,4) \tilde T_1 (2)
+ T_3 (2,3,4) \tilde T_1 (1)
$$
$$+T_2 (1,4) \tilde T_2 (2,3) + T_2 (2,4) \tilde T_2 (1,3) + T_2 (3,4)
\tilde T_2 (1,2)+T_1(4) \tilde T_3 (1,2,3). \eqno{(3.1)}
$$
Note that we only consider terms with field operator $:A(2)A(4):$.
According to lemma 1, the first, the third and the fifth term in (3.1)
vanish in the adiabatic limit.
Furthermore, the second term $R'_{42}$ gives the same contribution
as $R'_{47}$, and the same holds true for $R'_{44}$ and $R'_{46}$.

\def\Li{\mbox{Li}_2}
\newcommand{\iu}{\int \limits_{-\infty}^{+\infty}}
\newcommand{\ip}{\int \limits_{0}^{\sqrt{p^2}}}

In x-space, the three-particle contribution $R'_{44}$ (Fig. 3a) is given by
$$
R'_{44} = r'_{44} (1,2,3,4) :A(2)A(4):, \eqno{(3.2)}
$$
$$
r'_{44}=2i D_m^{AF} (2-3) D_m^F (1-4) D_m^{+} (1-2)
D_m^{+} (4-3) D_{m_3}^{+} (1-3)  \eqno{(3.3)}
$$
The Fourier transform of (3.3) is $(y_i=x_i-x_4)$
$$
r'_{44}(p_1,p_2,p_3) = (2 \pi)^{-6} \int dy_1dy_2dy_3 \, r'_{44}
(x_1,x_2,x_3,x_4) \, e^{ip_1y_1+ip_2y_2+ip_3y_3}
$$
$$
=2(2 \pi)^{-11} \int dq dp' \frac{1}{(p_2+q)^2 -m^2-i0} \, \frac{1}
{(p_1-q-p')^2-m^2+i0} \Theta(q^0) \delta(q^2-m^2)
$$
$$
\Theta(-p_2^0-p_3^0-q^0-{p'}^0) \delta((
p_2+p_3+q+p')^2 -m^2) \Theta({p'}^0) \delta({p'}^2-m_3^2)
\eqno{(3.4)}
$$
which becomes in the adiabatic limit $p_1,p_3 \rightarrow 0 \, , \, p_2=p$
$$
r'_{44}(p)=2(2 \pi)^{-11} \int dq dp' \frac{1}{(p+q)^2-m^2-i0} \,
\frac{1}{(q-p-p')^2-m^2+i0}
$$
$$
\Theta(q_0) \delta(q^2-m^2) \Theta({p'}_0-q_0) \delta((p'-q)^2-m^2)
\Theta(-p_0-{p'}_0) \delta((p+p')^2-m_3^2). \eqno{(3.5)}
$$

In fact, the calculation of (3.5) has already been performed in [2]
by G.K\"all\'en and A.Sabry. Our calculations confirm their result exactly,
namely:
$$
r'_{44}(p) = \frac{1}{4(2\pi)^9}\frac{1}{p^2} \Theta (-p_0) \Theta(p^2-4m^2)
\, B(z), \eqno{(3.6)}
$$
$$
B(z)= 3 \Li (z)+2 \Li (-z)+\log z \log (1-z) +\log z \log (1+z)
+ \frac{1}{4} \log^2 z - \frac {\pi^2}{3} + \log z \log \frac{m_3}{m},
\eqno{(3.7)}
$$
up to terms that vanish for $m_3 \rightarrow 0$.
Here we have introduced the well-known dilogarithm $\Li$,
and the variable
$$
z= \frac{1-\sqrt{1-4m^2/p^2}}{1+\sqrt{1-4m^2/p^2}}, \eqno{(3.8)}
$$
which varies from $0$ to $1$ for $p^2 \in [4m^2,\infty)$.

The two-particle contribution $r'_{42}$ (Fig.3b) is
$$
r'_{42}(p) = 2i (2 \pi)^{-2} \int dp' \, D_{m}^{+}(-p') \Lambda_3 (p',p'-p)
D_{m}^{+} (p'-p). \eqno{(3.9)}
$$
Here, $\Lambda_3$ is the vertex function. Since its first
argument lies in the backward light-cone and the second one in the
forward light-cone, we are in the case of lower signs in (2.13), because of
our convention (2.27). Then the retarded vertex function
$\Lambda_3^{ret}$ is given by the following dispersion integral:
$$
\Lambda_3^{ret} (p,q) = -\frac{i}{2 \pi} \iu dt \,
\frac{d_3 (tp,tq)}{1-t-i0}. \eqno{(3.10)}
$$
Again, the computation of (the real part of) $r'_{42}$ can be carried out in a
straightforward manner.
As we know from (2.30), the vertex function consists of
three different parts. The first one leads to an infrared finite
contribution to $r'_{42}$:
$$
{r'}^{1}_{42} (p) = \frac{1}{4 (2 \pi)^9} \frac{1}{p^2} \Theta(-p_0) \Theta
(p^2-4m^2) \, C_1(z), \eqno{(3.11)}
$$
$$
C_1(z) = 2 \Li (z)+2 \log z \log (1-z)- \frac{1}{2} \log^2 z+\frac{\pi^2}{6},
 \eqno{(3.12)}
$$
whereas the second and third term are equal and contain an infrared divergent
term $\sim \log (m_3/m)$, which cancels the infrared divergence in (3.7):

$$
{r'}^{2}_{42} (p) = \frac{1}{4 (2 \pi)^9} \frac{1}{p^2} \Theta(-p_0) \Theta
(p^2-4m^2) \, C_2(z), \eqno{(3.11)}
$$
$$
C_2(z) = - \Li (z)- \log z \log (1-z)+ \frac{1}{4} \log^2 z+\frac{\pi^2}{6}
-\log z \log \frac{m_3}{m}.  \eqno{(3.12)}
$$

Now $d_4=r'_4-a'_4$ can immediately be written down if we note that
$$
r'_{4i}(p) = a'_{4i}(-p), \quad i=1\ldots,7. \eqno{(3.13)}
$$

\section{The Splitting of $\bf{d_4(p)}$}
After having calculated the causal distribution $d_4$
$$
d_4 (p) = -\frac{1}{2 (2 \pi)^9} \frac{1}{p^2}
\mbox{sgn} \, p_0 \, \Theta(p^2-4m^2) \, J(z),
$$
$$
J(z)= 4\Li (z)+2 \Li (-z) + 2 \log(z) \log (1-z) + \log(z) \log(1+z),
\eqno{(4.1)}
$$
we must decompose it into retarded and advanced parts.
The retarded distribution $r_4$ is given in the forward light-cone
according to (2.13)
$$
r_4 (p) = \frac{i}{2 \pi} \int \limits_{-\infty}^{+\infty} dt
\frac{d_4(tp)}{1-t+i0}
$$
$$
= -\frac{i}{2 \pi} \frac{1}{2(2 \pi)^9}
\frac{1}{p^2} \int \limits_{-\infty}^{+\infty}
\frac{\mbox{sgn} \, t \, \Theta(t^2p^2-4m^2)}{t^2(1-t+i0)} J \Biggl( \frac
{\sqrt{t^2p^2}-\sqrt{t^2p^2-4m^2}}{\sqrt{t^2p^2}+\sqrt{t^2p^2-4m^2}}\Biggr)
\, dt. \eqno{(4.2)}
$$
It is very convenient to introduce the new integration variable
$$
x =
\frac{\sqrt{t^2p^2}-\sqrt{t^2p^2-4m^2}}{\sqrt{t^2p^2}+\sqrt{t^2p^2-4m^2}},
\quad t= \frac{1}{\sqrt{b}}
\frac{x+1}{\sqrt{x}}, \quad b=p^2/m^2, \quad
\frac{dt}{dx}= \frac{1}{\sqrt{b}} \frac{x-1}{2x \sqrt{x}}. \eqno{(4.3)}
$$
This leads to ($p \in V^+$)
$$
r_4 (p) = -\frac{i}{2(2 \pi)^{10}p^2}
\int \limits_{0}^{1} dx \Bigl\{ -\frac{2}{x+1}
+\frac{1}{x-z} + \frac{1}{x-1/z} \Bigr\}  \, J(x). \eqno{(4.4)}
$$
The imaginary part $i0$ in the denominator of (4.2) is included in $p^2$ as
discussed at the end of this section.

From now on, $r_4 (p)$ is considered as a function of z (3.8)
$$
z= \frac{1}{2} \Bigl(b-2-b \sqrt{1-4/b} \Bigr)
=  \frac{1-\sqrt{1-4m^2/p^2}}{1+\sqrt{1-4m^2/p^2}}. \eqno{(4.5)}
$$
Then we note that $z \in [0,1]$ for $p^2 \in [4m^2, \infty)$, and
$$
R(z)=\int \limits_{0}^{1} dx \Bigl\{ -\frac{2}{x+1} + \frac{1}{x-1/z}
+\frac{1}{x-z} \Bigr\} J(x) \eqno{(4.6)}
$$
has the property $R(z) = R(1/z)$.
The integral
$$
R_1= -2 \int \limits_{0}^{1} \frac{dx}{x+1} J(x) = - \frac{9}{4} \zeta(3)
\eqno{(4.7)}
$$
is just a constant and can be calculated with the formulae given in [10,11].
The calculation of
$$
R_2 (z) = \int \limits_{0}^{1}\frac{J(x)}{x-z}
+  \int \limits_{0}^{1}\frac{J(x)}{x-1/z}
=R_3 (z) + R_4 (z) \eqno{(4.8)}
$$
is most easily performed by first calculating the derivatives
of $R_3$ and $R_4$:
$$
R'_3 (z) = \partial_z \Bigl\{\frac{1}{z} \int \limits_0^1 dx \, \frac{J(x)}
{x/z-1} \Bigr\} = \partial_z \Bigl\{\int \limits_0^{1/z} dt \,
\frac{J(tz)}{t-1} \Bigr\}
= \frac{J(1)}{z(z-1)} + \frac{1}{z} \int \limits_0^1 dx \frac{x}{x-z} {J'}(x),
$$
$$
R'_4(z) = \frac{J(1)}{z-1} + \int \limits_0^1 dx \, \frac{J'(x)}{1/x-z}.
\eqno{(4.9)}
$$
This is a helpful trick, but all integrals coming up in (4.6) are also
discussed in the literature mentioned above.
We give the separate results for the two-particle and three-particle
expressions:
$$
\int \limits_{0}^{1} dx \Bigl\{ -\frac{2}{x+1} + \frac{1}{x-1/z}
+\frac{1}{x-z} \Bigr\} B(x)
$$
$$
=5 \mbox{Li}_3 (z) + 3 \mbox{Li}_3 (-z) -3 \mbox{Li} (z) \log z
-2 \mbox{Li} (-z) \log z -\frac{1}{2} \log^2 z \log (1-z)
$$
$$
-\frac{1}{2} \log^2 z \log (1+z) - \frac{1}{12} \log^3 z
+\frac{\pi^2}{2} \log (1+z) + \frac{\pi^2}{2} \log z  + \frac{3}{4} \zeta (3)
+ \frac{\pi^2}{2} \log 2 , \eqno{(4.10)}
$$
and
$$
\int \limits_{0}^{1} dx \Bigl\{ -\frac{2}{x+1} + \frac{1}{x-1/z}
+\frac{1}{x-z} \Bigr\} C(x)
$$
$$
=\mbox{Li}_3 (z) - \mbox{Li} (z) \log z - \frac{1}{2} \log^2 z \log (1-z)
+ \frac{1}{12} \log^3 z
$$
$$
+\pi^2 \log(1-z) - \frac{\pi^2}{2} \log z + \frac{3}{4} \zeta (3)
- \frac{\pi^2}{2} \log 2, \eqno{(4.11)}
$$
where $C(x)=C_1(x)+C_2(x)$.
Here we have introduced the trilogarithm, which is defined by
$$
\mbox{Li}_3 (z) = \int \limits_0^z dx \frac{\mbox{Li}(x)}{x}. \eqno{(4.12)}
$$
The infrared divergent terms in $B$ and $C$ which cancel mutually in $r_4$
have already been omitted. Finally $r_4$ is given by
$$
r_4(z) =- \frac{i}{2(2 \pi)^{10} p^2} R(z), \eqno{(4.13)}
$$
\def\L3{\mbox{Li}_3}
$$
R(z) = 6 \L3 (z) + 3 \L3 (-z) - 4 \Li (z) \log (-z) -2 \Li (-z) \log (-z)
$$
$$
-\log^2 (-z) \log (1-z)
- \frac{1}{2} \log^2 (-z) \log (1+z) + \frac{3}{2} \zeta(3). \eqno{(4.14)}
$$
If $R(z)$ becomes complex the sign of the imaginary parts is determined by
the $i0$ in (4.2) as follows. For arbitrary time-like
$p$, the $i0$ goes over into $i0p_0$ (see (2.13) and [2], Chap.3.6).
To obtain the full time-ordered distribution $t_4 (p)$ we have to subtract
$r'_4 (p)$. This changes $i0p_0$ into $i0$ again. Consequently, $t_4 (p)$ is
given by (4.13) with $p^2$ substituted by $p^2+i0$, i.e. $z \rightarrow z+i0$.
This fixes the signs of the imaginary parts at the logarithmic cuts in (4.14).
For space-like $p$, $t_4 (p)$ is simply given by (4.13) because $r'_4 (p)$
now vanishes and $R(z)$ is real.

\section{The Electron Propagator in Fourth Order}
After having demonstrated the instructive example of vacuum polarization,
we proceed now with the discussion of the electron propagator in fourth order
(Fig.1b).
Again, we start from the general expression (3.1)
and consider only terms with field operator $:\varphi(2) \varphi^+ (4):$.
Just as in the case of vacuum polarization, the first, the third and the fifth
term in (3.1) vanish in the adiabatic limit, and the second and the
seventh term
$$R'_{42}=T_3 (1,3,4) \tilde T_1 (2), \quad R'_{47}=T_1(4) \tilde T_3 (1,2,3)
\eqno{(5.1)}
$$ give the same contribution to the propagator.
But we have to distinguish the different cuts in $R'_{44}$ and $R'_{46}$.

In x-space, the three-particle contribution in $R'_{46}$ with
two photons and one electron as intermediate state (Fig.4) is given by
$$
{R'}_{46}^{1} = {r'}_{46}^{1} (1,2,3,4) :\varphi(2)\varphi^+(4):, \eqno{(5.2)}
$$
$$
{r'}_{46}^{1}=i D_m^F (3-4) D_m^{AF} (1-2) D_m^{+} (3-1)
D_{m_3}^{+} (4-1) D_{m_3}^{+} (3-2) \eqno{(5.3)}
$$
The Fourier transform is after the adiabatic limit
$$
{r'}_{46}^{1} (p)=(2 \pi)^{-11} \int dq dp' \frac{1}{(p+q)^2-m^2-i0} \,
\frac{1}{(q-p-p')^2-m^2+i0}
$$
$$
\Theta(q_0) \delta(q^2-m_3^2) \Theta({p'}_0-q_0) \delta((p'-q)^2-m_3^2)
\Theta(-p_0-{p'}_0) \delta((p+p')^2-m^2). \eqno{(5.4)}
$$
Straightforward calculation of (5.4) leads to the simple result
$$
{r'}_{46}^{1} = \frac{1}{8(2 \pi)^9}\frac{1}{p^2} \Theta(-p_0)
\Theta(p^2-m^2) B(x^2), \eqno{(5.5)}
$$
$$
B(x^2)= \frac{1}{2} \Li (x^2) + \frac{1}{2} \log(x^2) \log(x^2-1) -
\frac{\pi^2}{12} \quad , \quad x^2 = p^2/m^2. \eqno{(5.6)}
$$
The two-particle contribution can be evaluated without any problem
with the aid of (2.30). We therefore quote only the result:
$$
{R'}_{42} = {r'}_{42} (1,2,3,4) :\varphi(2)\varphi^+(4):, \eqno{(5.7)}
$$
$$
{r'}_{42} (p) = \frac{1}{4(2 \pi)^9}\frac{1}{p^2} \Theta(-p_0)
\Theta(p^2-m^2) C(x^2), \eqno{(5.8)}
$$
$$
C(x^2)=\frac{1}{2} \Li (x^2) - \frac{\pi^2}{12}. \eqno{(5.9)}
$$
In (5.6) and (5.9) only the real parts of the dilogarithms contribute.

Since we have to treat the three-particle contribution with three electrons
as intermediate state (${r'}_{44}^2$ and ${r'}_{46}^2$) separately, we
already give the splitting results for the expressions obtained so far. The
splitting procedure leads to the same type of integrals as in the case
of vacuum polarization in Sect.4, hence we refer again to [10,11].
We have for $p$ in the forward light-cone and $p^2 > m^2$
$$
d_1(p)=-\frac{1}{8(2 \pi)^9} \frac{1}{p^2} \mbox{sgn} \, p_0 \,
\Theta(p^2-m^2)\Bigl[3 \Li (x^2) + \log x^2 \log(x^2-1)
-\frac{\pi^2}{2} \Bigr], \eqno{(5.10)}
$$
$$
r_1(p)=\frac{i}{2\pi} \int \limits_{-\infty}^{+\infty} dt
\frac{d_1 (tp)}{1-t+i0} \eqno{(5.11)}
$$
$$
=-\frac{i}{8(2 \pi)^{10} p^2}\Bigl[4 \mbox{Li}_3 (1-x^2) - 3 \log(1-x^2) \Li
(1-x^2)-\log x^2 \log^2 (1-x^2) - 4 \zeta(3) \Bigr].\eqno{(5.12)}
$$

The calculation of the second diagram in Fig.4
$$
{r'}_{44}^{2} (p)=(2 \pi)^{-11} \int dq dp' \frac{1}{(p+q)^2-m_3^2-i0} \,
\frac{1}{(q-p-p')^2-m_3^2+i0}
$$
$$
\Theta(q_0) \delta(q^2-m^2) \Theta({p'}_0-q_0) \delta((p'-q)^2-m^2)
\Theta(-p_0-{p'}_0) \delta((p+p')^2-m^2). \eqno{(5.13)}
$$
is difficult and requires extra consideration.
${r'}_{44}^2 (p)$ turns out to be infrared finite, so we may drop the
small photon mass in the following.
Making use of all $\delta$-distributions, the integral
$$
I=\int dq \Theta(q_0) \delta(q^2-m^2) \Theta(p'_0-q_0) \delta((p'-q)^2-m^2)
\frac{1}{(q+p)^2-i0} \, \frac{1}{(q-p'-p)^2+i0} \eqno{(5.14)}
$$
can be transformed into
$$
I= -\frac{\pi}{2} \frac{\Theta(p'_0) \Theta({p'}^2-4m^2)}
{\sqrt{(p' \tilde p)^2 -{p'}^2 {\tilde p}^2}}
\int \limits_{x_1}^{x_2} dx \frac{1}{4(x+m^2)(x+p \tilde p)}, \eqno{(5.15)}
$$
where
$$
\tilde p = -p-p' \quad , \quad x_{1,2} = \frac{1}{2} p' \tilde p \mp
\frac{1}{2} \sqrt{(p' \tilde p)^2 -{p'}^2 {\tilde p}^2} \sqrt{1 -
\frac{4m^2}{{p'}^2}}. \eqno{(5.16)}
$$
Then we obtain
$$
{r'}_{44}^2 (p) = -\frac{1}{4(2 \pi)^9} \frac{1}{p^2} \Theta(-p_0)
\Theta(p^2-9m^2) \int \limits_{\sqrt{p^2 m^2}}
^{\frac{1}{2}(p^2-3m^2)} dy \int \limits_{x_1}^{x_2} dx \frac{1}
{4(x+m^2)(x-y)}, \eqno{(5.17)}
$$
$$
x_{1,2} = \frac{1}{2}(y-m^2) \mp \frac{\xi}{2}, \quad
\xi=(y^2-p^2m^2) \sqrt{1-\frac{4m^2}{p^2+m^2-2y}}. \eqno{(5.18)}
$$
Introducing $x=(y+z-m^2)/2$, $d_2(p) = 2({r'}_{44}^2(p)-{a'}_{44}^2(p))$
becomes
$$
d_2 (p) = -\frac{1}{4(2 \pi)^9}\frac{1}{p^2}
\mbox{sgn}\,p_0 \, \Theta (p^2-9m^2) \int
\limits_{\sqrt{p^2m^2}}^{\frac{1}{2}(p^2-3m^2)} dy \int \limits_{-\xi}^
{+\xi} dz\frac{1}{(y+z+m^2)(y-z+m^2)} \eqno{(5.19)}
$$

At this stage, we will not proceed the same way as in the case of vacuum
polarization. We first apply
the splitting formula to (5.19)
$$
r_2 (p) = \frac{i}{2 \pi} \int \limits_{-\infty}^{+\infty} dt
\frac{d_4^2(p)}{1-t+i0}
$$
$$
= -\frac{i}{4(2 \pi)^{10} p^2} \int \limits_{9\frac{m^2}{p^2}}^
{+\infty} \frac{ds}{s(1-s+i0)}
\int \limits_{\sqrt{p^2m^2s}}^{\frac{1}{2}(p^2s-3m^2)} dy \int
\limits_{-\xi(s,y)}^
{+\xi(s,y)} dz\frac{1}{(y+z+m^2)(y-z+m^2)}, \eqno{(5.20)}
$$
$$
\xi(s,y) = \sqrt{y^2-p^2m^2s} \sqrt{1+\frac{4m^2}{2y-m^2-p^2s}}, \quad
p \in V^+ \eqno{(5.21)}
$$
and perform a partial integration with respect to $s$.
This leads to the following integral:
$$
r_2 (p) = \frac{i}{2(2 \pi)^{10} p^2} \int
\limits_{9\frac{m^2}{p^2}}^{+\infty} ds \log \Bigl( \frac{s}{1-s-i0} \Bigr)
\int \limits_{\sqrt{p^2m^2s}}^{\frac{1}{2}(p^2s-3m^2)} dy
\frac{\frac{d}{ds} \xi(s,y)}{(y+m^2)^2-\xi^2(s,y)} \eqno{(5.22)}
$$
$$
= \frac{i}{2(2 \pi)^{10} p^2} \int \limits_{9/x^2}^
{+\infty} ds \log \Bigl( \frac{s}{1-s-i0} \Bigr)
\int \limits_{\sqrt{x^2s}}^{y_2} dy
\frac{\xi'}{(y+1)^2-\xi^2}, \eqno{(5.23)}
$$
$$
x^2=p^2/m^2 \quad , \quad y_1 = \frac{1}{2}(x^2 s+1), \quad
\quad y_2 = \frac{1}{2}(x^2 s-3) \quad ,
$$
$$
\xi = \sqrt{y^2-x^2s} \sqrt{\frac{y_2-y}{y_1-y}}. \eqno{(5.24)}
$$
We obtain after some simple manipulations
$$
r_2= \frac{i}{2(2\pi)^{10} m^2} \int \limits_{9/x^2}^{\infty}
ds \frac{\log \Bigl( \frac{s}{1-s-i0} \Bigr)}{(x^2s-1)^2}
$$
$$
\int \limits_{\sqrt{x^2s}}^{y_2} \frac{dy}{\sqrt{(y^2-x^2s)(y-y_1)(y-y_2)}}
\Bigl[ (x^2s-1)y-\frac{1}{4}(x^2s+1)^2+1 \Bigr]. \eqno{(5.25)}
$$

The last integral can be expressed by complete elliptic integrals of the first
and third kind. It is
$$
\int \limits_{\sqrt{x^2s}}^{y_2} \frac{dy}{\sqrt{(y^2-x^2s)(y_1-y)(y_2-y)}}
= \frac{2}{\sqrt{(y_1-\sqrt{x^2s})(y_2+\sqrt{x^2s})}} \mbox{K} (k),
\eqno{(5.26)}
$$
where
$$
k^2 = \frac{(y_2-\sqrt{x^2s})(y_1+\sqrt{x^2s})}{(y_1-\sqrt{x^2s})(y_2+\sqrt
{x^2s})}, \eqno{(5.27)}
$$
and
$$
\int \limits_{\sqrt{x^2s}}^{y_2} \frac{ydy}{\sqrt{(y^2-x^2s)(y_1-y)(y_2-y)}}=
$$
$$
\frac{2}{\sqrt{(y_1-\sqrt{x^2s})(y_2+\sqrt{x^2s})}} \Bigl[ 2 \sqrt{x^2s}
\Pi(\alpha^2,k) - \sqrt{x^2s} \mbox{K} (k) \Bigr], \eqno{(5.28)}
$$
$$
\alpha^2=\frac{y_2-\sqrt{x^2s}}{y_2+\sqrt{x^2s}}. \eqno{(5.29)}
$$
Introducing $\lambda=\sqrt{x^2s} \in [3,\infty)$,
we find that the modulus $k$ and the
parameter $\alpha^2$ are related by the following identities:
$$
\alpha^2=\frac{(\lambda-3)(\lambda+1)}{(\lambda+3)(\lambda-1)} \quad , \quad
k^2=\frac{(\lambda-3)(\lambda+1)^3}{(\lambda+3)(\lambda-1)^3}. \eqno{(5.30)}
$$
Then it is in fact possible to express $\Pi(\alpha^2,k)$ by $\mbox{K}(k)$:
$$
\Pi(\alpha^2,k)=\frac{\lambda+3}{6} \mbox{K}(k). \eqno{(5.31)}
$$
This has already been observed by A.Sabry [12]. Our relation follows from his
equation (85) by the substitution $\lambda \rightarrow 1/\lambda$.
Finally we arrive at a remarkably simple expression for $r_2$:
$$
r_2 (x)= \frac{i}{6(2\pi)^{10}m^2} \int \limits_{9/x^2}^{\infty} ds
\frac{\lambda-3}{(\lambda-1)^2(\lambda+1)} \sqrt{\frac{\lambda+3}{\lambda-1}}
\log \Bigl( \frac{s}{1-s-i0} \Bigr) \mbox{K}(k). \eqno{(5.32)}
$$

So far we have calculated the retarded distribution for time-like
$x^2=p^2/m^2$.
By the same argument as given at the end of the last section, this also gives
the time-ordered distribution $t_2 (p)$ for arbitrary $p^2$. To get the full
fermion propagator, one must add $t_1 (p)$ given by (5.12) with
$x^2 \rightarrow x^2 + i0$. The integral in (5.32), which can easily
be computed numerically, is in agreement with eq. (30) of Broadhurst [4].
Our funny normalization factors are the correct ones for the calculation of
S-matrix elements according to (2.1).

\vskip 1cm
{\it References}\vskip 1cm

[1] H. Epstein, V. Glaser, Ann.Inst.Poincar\'e A 29 (1973) 211

[2] G.Scharf, Finite quantum electrodynamics, the causal approach,

\hskip 0.5cm (second edition, Springer, Berlin Heidelberg New York, 1995)

[3] G.K\"all\'en, A.Sabry, Dan.Mat.Fys.Medd. 19, No.17 (1955) 1

[4] D.J.Broadhurst, Z.Phys. C 47 (1990) 115

[5] J.A.Mignaco, E.Remiddi, Nuov.Cim. 60 A (1969) 519

[6] R.Barbieri, J.A.Mignaco, E.Remiddi, Nuov.Cim. 11 A (1972) 824, 865

[7] M.J.Levin, E.Remiddi, R.Roskies, QED, ed. T.Kinoshita,

\hskip 0.5cm (World Scientific, 1990) p.162-321

[8] G.'t Hooft, B.Veltman, Nucl.Phys. B 153 (1979) 365

[9] G.Weiglein, R.Scharf, M.B\"ohm, Nucl.Phys. B 416 (1994) 606

[10] L.Levin, Polylogarithms and associated functions (North Holland, New York,
1981)

[11] A.Devoto, D.W.Duke, Riv.Nuov.Cim. 7 (1984) 1

[12] A.Sabry, Nucl.Phys. 33 (1962) 401

\newpage

\vskip 1cm
{\it Figure Captions}\vskip 1cm

Fig.1 The two-loop master diagrams:

\hskip 1.0cm (a) the 'photon' propagator;

\hskip 1.0cm (b) the 'electron' propagator

Fig.2 The vertex diagram with arbitrary masses

Fig.3 (a) The three-particle cut, (b) the two-particle cut for fourth order

\hskip 1.0cm vacuum polarization

Fig.4 The two different three-particle cuts in the electron propagator


\end{document}